\documentclass[prl,twocolumn,preprintnumbers,reprint]{revtex4-1} 
\usepackage{dcolumn,bm}
\usepackage{amsmath,amssymb,amsfonts,graphics,graphicx,amscd,amsfonts,color}
\usepackage[colorlinks=true,allcolors=blue]{hyperref}
\usepackage{subfigure}
\newcommand{\dS}{dS }
\newcommand{\CC}{$\Lambda\,$}
\newcommand{\RI}{\textbf{$I$} }
\newcommand{\RII}{\textbf{$II$} }
\newcommand{\RIII}{\textbf{$III$} }
\newcommand{\RIV}{\textbf{$IV$} }
\newcommand{\RV}{\textbf{$V$} }

\newcommand{\ein}{\emph{In} }
\newcommand{\eout}{\emph{Out} }
\newcommand{\dOmt}{\,d\Omega^{2}_{\it 2}}

\newcommand{\bwt}{\begin{widetext}}
\newcommand{\ewt}{\end{widetext}}

\date{\today}

\begin{document}

\title{de Sitter as a Resonance}

\author{Jonathan Maltz$^{a,b}$} 
\author{Leonard Susskind$^{b}$}
\affiliation{${}^a$ Center for Theoretical Physics and Department of Physics, University of California at Berkeley, Berkeley, California, 94720,USA}
\affiliation{
${}^b$ Stanford Institute for Theoretical Physics, Stanford University, Stanford, California 94305, USA}

\preprint{SU-ITP-16/18}

\begin{abstract}
A quantum mechanical formulation of de Sitter  cosmological spacetimes still eludes string theory.  In this paper we conjecture a potentially rigorous framework in which the status of de Sitter space is the same as that of a  resonance in a scattering process. We conjecture that transition amplitudes between certain states with asymptotically supersymmetric flat vacua contain resonant poles characteristic of metastable intermediate states. A calculation employing constrained instantons illustrates this idea.
\end{abstract}

\maketitle
\section{Introduction}

The Universe is not just expanding; the expansion is accelerating \cite{Riess:1998cb,Perlmutter:1998np,Spergel:2003cb,Smoot:1992td,Perlmutter:1998np,Fixsen:1996nj}.  This implies that our Universe is asymptotically de Sitter (dS) \cite{Nagamine:2002wi,Busha:2003sz,Dodelson:1282338,Frieman:2008sn,weinberg2008cosmology}.
The \dS metric expressed in global coordinates is

\begin{equation}
ds^2=-dt^2+l^{2}_{\text{ds}}\cosh^{2}{\Big[\sqrt{\frac{t}{l_{\text{ds}}}}\Big]}\big(d\psi^{2}+\sin^{2}{\psi}\dOmt\big)
\end{equation}   with $l_{dS} = \sqrt{\frac{3}{\Lambda}}$. Expressed in conformal time
\begin{equation}
ds^2_{dS} = \frac{3}{\Lambda \cos^{2}{\eta}}\Big\{-d\eta^2 + d\psi^2 + \sin^{2}{\psi}\dOmt\Big\}
\end{equation} 
where 
\begin{equation}\label{relation}
\cosh{\Big[\sqrt{\frac{\Lambda}{3}}t\Big]}=\frac{1}{\cos{\eta}}
\end{equation} 

Our current understanding of quantum gravity is dependent on the existence of stable asymptotically cold boundaries, and correlation functions evaluated at those boundaries. Supersymmetric anti de Sitter space (AdS) and asymptotically Minkowski space are examples. Boundary correlators and \emph{S}-matrix elements are the mathematical objects that the theory is built out of  \cite{Green:1987sp,Green:1987mn,Polchinski:1998rq,Polchinski:1998rr,Aharony:1999ti}. No such understanding exists for de Sitter space.

Even if de Sitter space were stable with respect to decay, its future spacelike asymptotic boundary would fluctuate, necessitating an integration over the geometry of future infinity---a problem that could be as complicated as any quantum gravity problem. In addition it is expected that all de Sitter vacua are unstable with respect to Coleman De Luccia (CDL) decay to flat and AdS vacua \cite{Dyson:2002nt,Susskind:2003kw,Dyson:2002nt}. This would further complicate the asymptotic future boundary, turning it into a superposition of fractals populated by crunches and hats.

But the existence of hats--- Friedmann-Robertson-Walker (FRW) patches with vanishing cosmological constant---creates a new opportunity for a rigorous framework for de Sitter space, in which it appears as a metastable state in a transition amplitude between two asymptotically flat states. Our purpose in this Letter is to define such a transition amplitude and show that resonant poles, associated with de Sitter intermediate states,  exist in its spectral representation. 

To illustrate the idea we begin with a configuration that resembles a time-symmetric slice of a de Sitter vacuum. The state can be propagated forward and backward in time to give past and future quantum superpositions of fractal boundaries, each containing an infinite number of hats. We make a gauge choice by picking a hat in the past and a hat in the future and transform them to the center of a causal diamond. The causal diamond of the hats comprise a universe for an observer who begins and ends in the past and future hats.

The complete details of the computation will be presented in a technical paper that one of us will publish concurrently \cite{Maltz:2016max}. Suggestions that \dS might be viewed as a resonance have occurred previously \cite{Freivogel:2004rd}; however, to our knowledge there has been  no calculation to establish this. We present one in this work.

\section{Resonances and the Causal Patch}

\begin{figure}[t]
\begin{center} 
\scalebox{0.9}{
\includegraphics[width=9cm]{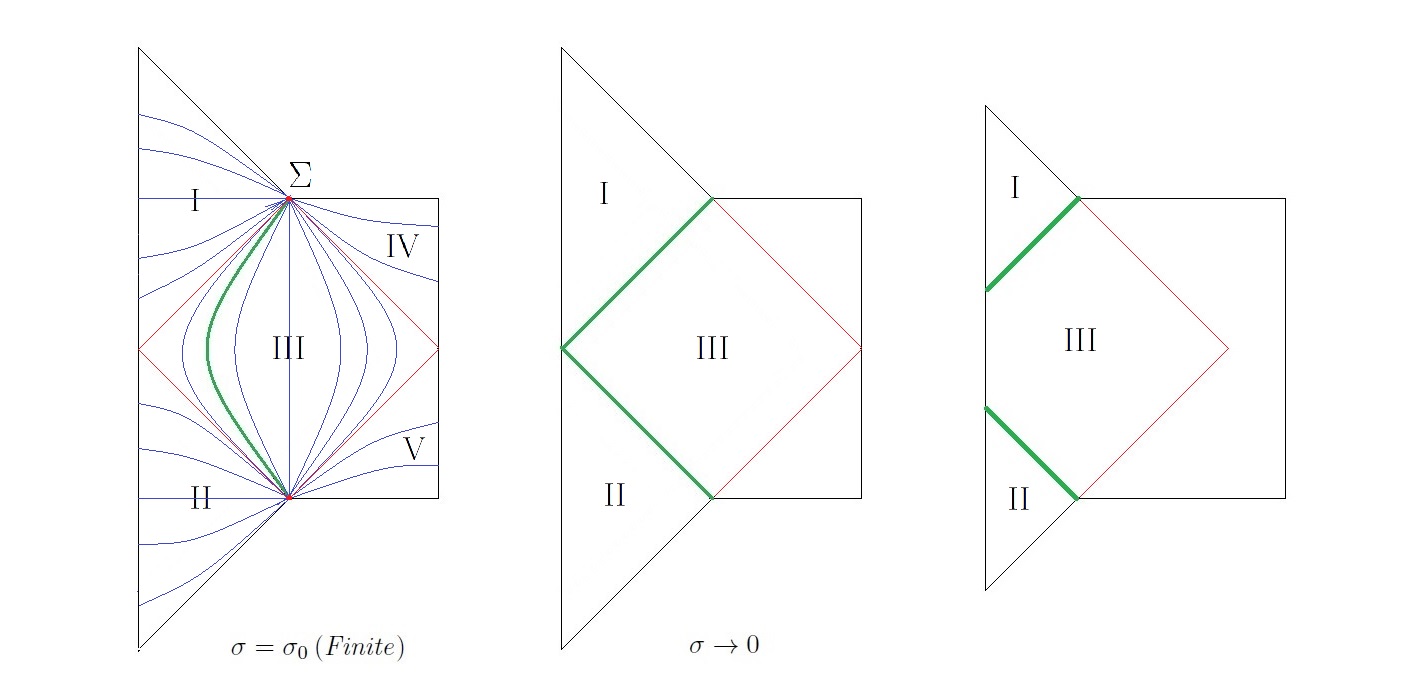}}
\end{center}
\caption{(Left) The Penrose diagram of the Lorentzian continuation of the CDL instanton solution \cite{PhysRevD.21.3305,Freivogel:2004rd,Freivogel:2006xu}. \RI and \RII are an open (k = -1) FRW Universes that are asymptotically flat \cite{PhysRevD.21.3305,Freivogel:2004rd,Freivogel:2006xu}. \RIV and \RV are asymptotically \dS. $\Sigma$ is the conformal 2-sphere defined by the intersection of the lightlike infinity of \RI and the spacelike infinity of \RIV. The curves indicate orbits of the $SO(3,1)$ symmetry, which acts as the conformal group on $\Sigma$ \cite{Freivogel:2006xu}. The red lines between \RIII, \RIV, and \RV represent the cosmological horizons in the \dS of the observer at $r=0$. The green curve in \RIII represents the domain wall between the FRW and \dS regions. (Center) The tensionless domain wall limit (right) the constrained CDL spacetime where the FRW bubbles are separated by $2\eta_0$, here \RIII is pure \dS and is the unregulated integration region.}
\label{fig:CDLconstrainedCDL}
\end{figure}

In \cite{Freivogel:2004rd} a simplified landscape of two minima, one  with $\Lambda > 0$ and the other with $\Lambda =0$, was considered and an $O(D-1)$ symmetric CDL instanton solution was worked out; the Penrose diagram for this spacetime is the left diagram of Fig. \ref{fig:CDLconstrainedCDL}.  On the spacelike slice in the middle of the left diagram of Fig. \ref{fig:CDLconstrainedCDL} a Hartle Hawking state for the spacetime can be constructed and evolved to an \eout state \footnote{ States must be restricted to those that do not possess enough energy and entropy to collapse the geometry into a singularity \cite{Banks:2002fe,Bousso:2006ge,Bousso:2010zi}. }. The information within the causal patch (regions \RI $+$ \RII $+$ \textbf{$III$}) is then all that is needed to capture all the information if horizon complementarity \cite{Susskind:1993if,Bigatti:1999dp,Banks:2001yp,Bousso:2002ju} is correct, as anything that passes out of the causal patch (goes into \textbf{$IV$}) will have a complementary description in terms of the highly scrambled Hawking radiation that will go into \textbf{$I$}. Therefore, a spacelike slice in \RI contains all the information from the Hartle-Hawking state and we can construct the \eout state there \footnote{This is an operating assumption of FRW/CFT were the spacelike slice is usually at late time.}.

A resonance is an intermediate metastable state  that can occur between any pair of initial and final states \footnote{A heuristic example is that many different  states (all those with enough mass-energy to form a BH) can lead to a dense collection of overlapping resonances: a BH.}. Many of these channels can be used to establish the existence of a resonance. We compute a spectral representation of the transition amplitude between \ein and \eout states, $\langle\text{\emph{Out}}|\text{\emph{In}}\rangle$, and show that it contains a pole characteristic of a \dS intermediate state \cite{goldberger2004collision,2008AIPC.1077...31R}. The location of the pole is a function of \CC. 

The transition is computed as  a path integral over all histories --- including all possible spacetime configurations, and field configurations --- that connect the \ein and \eout states.
A mathematically tractable, though not realistic channel, as it is entropically suppressed, is to construct the \ein and \eout states by evolving in a time reversal symmetric manner from a semiclassical spacelike slice in the middle of \textbf{$III$}. 
We are not proposing that this is the true cosmological history of our Universe; we are proposing that such a pole  in such an amplitude provides a precise definition within the context of supersymmetic backgrounds of a \dS space \footnote{Time reversal symmetric configurations while entropically suppressed are a useful way of constructing resonances \cite{goldberger2004collision}.}. This is the same logic that applies to any metastable state in quantum mechanics.

Consider the CDL instanton in the thin wall tensionless domain wall limit (see Center Diagram Fig.\ref{fig:CDLconstrainedCDL}), which can be constructed using Barrab\`es-Israel null junctions conditions \cite{PhysRevD.43.1129}. The details of the nucleation process are not important in what follows.

We define the amplitude as a path integral over the causal patch containing the hats. We do not try to justify this; we define such an amplitude to be the object of interest and show that it has a \dS pole in the spectral representation. This definition eliminates the complicated fractal boundaries in regions $IV$ and $V$.

An off shell continuation of this configuration has the nucleation point and its time reversal separated by a finite conformal time $2\eta_0$ (see the right diagram of Fig.\ref{fig:CDLconstrainedCDL}). In what follows we truncate the path integral to an integral over $\eta_0$, (\ref{calschematic}). This deformed spacetime is not a true instanton of the orginal CDL equations but has the status of a constrained instanton solution \cite{Frishman:1978xs,Affleck:1980mp,Nielsen:1999vq}, with the constraint that the separate FRW regions are separated by a given proper time along geodesic $\psi =0$. We refer to this spacetime as the \emph{constrained} CDL instanton. 
 The path integral over histories (\ref{calschematic}) contains deformations of this geometry including metric and field fluctuations about the instanton solution as well as nonperturbative effects such as further vacuum decay outside the hat.  This minisuperspace approximation allows  us to focus on the first contribution to this path integral over all histories of the metric and fields (\ref{calschematic}) by only integrating over histories where no particle content is excited and the integration $\eta_0$. Defining the proper time along $\psi=0$ to be $2 t_0$ using (\ref{relation}),
\begin{align}\label{calschematic}
&\langle \eout|\ein\rangle =\int\,\mathcal{D}g\,\mathcal{D}\varphi\,\exp{\{i\,S[g,\varphi]\}}\nonumber\\&\hspace{0.15in}\sim\mathcal{N}\int\,d\,t_0\exp{\{i\,S[t_0]\}}+\text{higher order terms.}
\end{align}

The higher order terms are weighted by powers of $l_{dS}$.
This expresses the amplitude as an integral over the relative time between the initial and final hats. The Fourier transform of the $t_0$ dependence defines the spectral representation of $\langle\text{\emph{Out}}|\text{\emph{In}}\rangle$.

\section{The Action In Liouville Gravity}\label{onepone}

\begin{figure}[t]
\begin{center}
\subfigure[ $\,1+1$ \emph{constrained} CDL spacetime]{\includegraphics[scale=0.25]{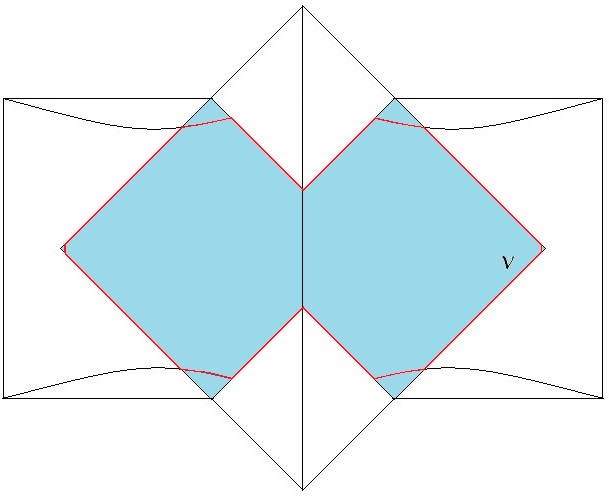}}\hspace{0.25in}
\subfigure[ $\,3+1$ \emph{constrained} CDL spacetime]{\includegraphics[scale=0.375]{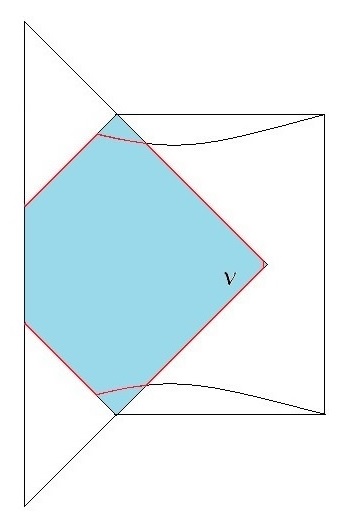}}
\end{center}
\caption{(a) The slices of constant  $r^{2}_0 = \frac{3\sin^{2}{\psi}}{\Lambda\cos^{2}{\eta}}$ are the curved surfaces intersecting the null lines at $\psi_{1}=\arctan{\Bigg[\frac{\cos{[\eta_{0}]}}{\sin{\eta_{0}}+\sqrt{\frac{3}{\Lambda r^{2}_{0}}}}\Bigg]}$ and $\psi_{2}=\frac{\pi}{2}-\arctan{\Bigg[\frac{\sin{\eta_{0}}-\sqrt{\frac{3}{\Lambda r^{2}_{0}}}}{\cos{\eta_0}}\Bigg]}$ as well as their reflection about $\psi=0$. The null domain walls dividing the \dS and FRW regions intersect $\psi=0$ at conformal time $\eta=\eta_0$ and $\eta=-\eta_0$.
(b) $\mathcal{V}$ for the $3+1$ spacetime.}\label{spacetime}
\end{figure}

In this section we illustrate the computation in the context of dS$_{2}$, which can be described by Lorentizan timelike Liouville quantum gravity \cite{Polchinski:1989fn,Polchinski:1990mh,Ginsparg:1993is,Harlow:2011ny}, which has $dS_{2}$ as a solution \cite{Ambjorn:1998fd,Loll:1999uu,Harlow:2011ny}. The action has bulk contributions and Gibbons-Hawking boundary contributions from the causal patch. In $1+1$ dimensions, the Gauss-Bonnet theorem implies that the boundary contributions integrate to the Euler characteristic of the geometry; therefore in $1+1$ dimensions we only need to consider the volume contribution to the action $S[\eta_0]$. In the approximations that we are employing the only contributions to the action $S[\eta_0]$ that are $\eta_0$ dependent are those of the purely \dS region \footnote{The domain wall stress tensor action contribution is $\eta_0$ independent as it is the energy density required to change the CC from $\Lambda$ to 0 in the transition region and does not depend on the position of the bubble.}, the shaded portion of  Fig. \ref{spacetime}(a). Gauge fixing the metric using conformal gauge $g_{\mu\nu}=e^{\phi_c}\eta_{\mu\nu}$ gives the Liouville action, $S_{L}[\eta_0] = -\frac{1}{16\pi b^{2}}\int\,d^{2}\xi\big(\eta^{ab}\partial_{a}\phi_c\partial_{b}\phi_c -16\lambda e^{\phi_c}\big)$ with $e^{\phi_c}=\frac{3}{\Lambda\cos^{2}{\eta}}$ and $\lambda =\pi\mu b^{2}$ so $\frac{1}{2\pi b^{2}}=\frac{\pi\mu}{2\pi\lambda}=\frac{4!\mu}{4\Lambda}$, \footnote{Here the timelike Liouville action is defined and evaluated via analytic continuation as shown in \cite{Harlow:2011ny} with the remaining $SL_{2}\mathbb{C}$ gauge redundancy fixed by the procedure of \cite{Maltz:2012zs}. $\mu$ is the Liouville CC and the ratio $\Lambda/\mu$ should be thought of as expressing $\Lambda$ in terms of the UV Planck scale.}.  The action must be regulated as it is IR divergent due to the infinite volume in the tips near future and past infinity. The regulator must be one that respects the \dS and Lorentz boost  symmetries \footnote{The authors would like to thank Ying Zhao for her many discussions and comments on this point.} in order to separate the divergence in an invariant way. Surfaces of constant $r^{2}_0= \frac{3\sin^{2}{\psi}}{\Lambda\cos^{2}{\eta}}$ serve as a cutoff. (In higher dimensions these are surfaces of constant transverse $\mathbb{S}^{d-1}$ radius in $d+1$ dimensions.)  Spacetime points move along these surfaces under boosts and rotations
. The regulated integration region is the section bounded by the red curve in Fig. \ref{spacetime}(a), $\mathcal{V}$.

Evaluating $-\frac{1}{16\pi b^{2}}\int_{\mathcal{V}}\,d^{2}\xi\{\eta^{ab}\partial_{a}\phi_c\partial_{b}\phi_{c}-16\lambda\,e^{\phi_c}\}=\frac{1}{4\pi b^{2}}\int_{\mathcal{V}}\,d\psi\,d\eta\,\frac{1+\sin^{2}{\eta}}{\cos^{2}{\eta}}$, 
and expanding it in a Laurent expansion in $w_0=1/r_0$ up to $O[w_0]$, yields

\begin{equation}\label{Vcutoff1p1}
S_{L}=\frac{-4\log{\Big|\sqrt{\frac{3}{\Lambda}}w_0\Big|}+4-\frac{\pi^{2}}{4}+\frac{\eta^{2}_{0}}{2}-2\log{\cos{\eta_0}}}{\pi b^{2}},
\end{equation} with $\frac{1}{\pi b^{2}}=\frac{4!\mu}{2\Lambda}$. The $-\frac{4!\mu}{2\Lambda}\big\{\log{\sqrt{\frac{3}{\Lambda}}w_0}-4+\frac{\pi^{2}}{4}\big\}$ term is the divergent contribution of the action that remains when $\eta_0=0$. It is just the action of the Lorentizan tensionless domain wall CDL instanton, $S_0$, in this limit, and when exponentiated can be absorbed into the overall normalization factor of (\ref{calschematic}). Expressing (\ref{Vcutoff1p1}) in terms of proper time $t_0$ and defining  $\tilde{S}_{L}= S_{L} -S_{0}$, up to $O[w_0]$,
\begin{align}
\tilde{S}_{L}&=
\frac{4!\mu}{4\Lambda}\Bigg\{\arctan^{2}{\Bigg[\tanh{\Bigg[\sqrt{\frac{\Lambda}{3}}t_0\Bigg]}\Bigg]}\nonumber\\&\hspace{0.25in}+\log{\cosh{\Bigg[\sqrt{\frac{\Lambda}{3}}t_0\Bigg]}}\Bigg\}.
\end{align}

The only term that grows with $t_0$ is $\log{\cosh{\Big[\sqrt{\frac{\Lambda}{3}}t_{0}\Big]}}=\sqrt{\frac{\Lambda}{3}}t_{0}+\log{|1+e^{-2\sqrt{\frac{\Lambda}{3}}t_0}|}-\log{2}$; the others are bounded.
Thus we find that for large $t_0$, $\tilde{S}$ behaves like $\sqrt{\frac{\Lambda}{3}}t_{0}$. Treating the  bounded term as a perturbation and Fourier transforming with respect to $t_0$ yields

\begin{align}\label{1p1pole}
\int^{\infty}_{0}d\,t_{0}\,e^{i (\tilde{S}_L[t_0] - \omega\,t_0)}&=\int^{\infty}_{0}d\,t_{0}\, e^{i(2\mu\sqrt{\frac{3}{\lambda}}t_{0}-\omega t_0)}+\ldots\nonumber\\&= \frac{i}{\omega -2\mu\sqrt{\frac{3}{\Lambda}}} + \rho_{1}[\omega] +\ldots,
\end{align}
thus revealing a pole in the spectral representation. One notes  that $2\mu\sqrt{\frac{3}{\Lambda}}$ is the energy of the static patch of dS, we take the existence of this pole to be the indication of an intermediate \dS vacuum. 

This indicates that the \dS can be thought of as a resonance in a transition amplitude. 

The pole in (\ref{1p1pole}) occurs at a real value of $\omega$ but this is an approximation. When the metastable character of the \dS vacuum is accounted for the cosmological constant obtains a small imaginary part determined by the CDL decay rate. This shifts the pole by a slightly imaginary amount, which is standard in the analysis of resonances \cite{0201503972,2008AIPC.1077...31R}.
\section{The 4 dimensional case}\label{tpo}
Let us repeat this in $3+1$ dimensions using the general relativity (GR) limit of the spacetime \footnote{This can be the low-energy limit of a string theory compactification from ten dimensions where supergravity reduces to GR. The CDL instanton being $O(D-1)$ symmetric can be formulated in ten dimensions and the asymptotically flat regions can be ten or eleven dimensional as the moduli can roll decompactifying the small dimensions in the domain wall region \cite{Freivogel:2006xu}.}. The regulated cutoff region $\mathcal{V}$ is shown in Fig.\ref{spacetime}(b). $\mathcal{V}$ is bounded by spacelike and  null surfaces. Hence we must append to the Einstein-Hilbert action \cite{Einstein:1916cd} the Gibbons-Hawking-York boundary  term \cite{PhysRevLett.28.1082,PhysRevD.15.2752}. The boundary term has to be generalized to null surfaces and corners as in \cite{Parattu:2015gga,Parattu:2016trq,PhysRevD.47.3275,Brown:2015lvg,Lehner:2016vdi}\footnote{The authors thank Ying Zhao for her helpful conversations on the necessity and evaluation of the corner terms.}. 
\begin{align}\label{actionmaster}
S&=\int_{\mathcal{V}}\frac{d^{4}x}{2\kappa}\sqrt{-g}\big(R-2\Lambda\big)-\sum_{i=2,4,6}\int_{\partial\mathcal{V}_{i}}d^{3}x\frac{\sqrt{h_{(i)}}K_{(i)}}{\kappa}\nonumber\\&\hspace{0.2in}+\sum_{i=1,3,5,7}\frac{1}{2\kappa}\int_{\partial\mathcal{V}_{i}}\,d^{2}x\sqrt{q_{(i)}}\mathit{\Theta} +\sum^{5}_{j=1}S_{\text{corner},(j)}.
\end{align} The second to last term is the null surface contribution and the last term is the five corner contributions that depend on a product of their boost angle and the area of the $\mathbb{S}^{2}$ at that point \cite{PhysRevD.47.3275,Lehner:2016vdi}. The geometry is closely related to Wick rotations of those in \cite{Brown:2015bva,Brown:2015lvg,PhysRevD.47.3275,Lehner:2016vdi} where the analysis of the null and corner terms was carried out. More detailed arguments on these terms are given in \cite{Maltz:2016max}.

Evaluating (\ref{actionmaster}) to $O[w_0]$, with $w_0=1/r_0$ and as a function of $t_0$, gives 
\begin{align}\label{totsactiont}
S&=\frac{4\pi4!}{2\kappa\Lambda}\Bigg\{\frac{1}{2}\log{\cosh{\Bigg[\sqrt{\frac{\Lambda}{3}}t_0\Bigg]}}+\frac{1-\sinh^{2}{\Big[\sqrt{\frac{\Lambda}{3}}t_0\Big]}}{8\cosh^{2}{\Big[\sqrt{\frac{\Lambda}{3}}t_0\Big]}}\nonumber\\&\hspace{0.25in}+\Big\{\frac{1}{4}-\log{4}\Big\}\tanh^{2}{\Bigg[\sqrt{\frac{\Lambda}{3}}t_0\Bigg]}\Bigg\}+ S_{0},
\end{align}
with $S_{0} = \frac{4\pi4!}{2\kappa\Lambda}\Big\{\frac{\Lambda}{2w^{2}_0}+\frac{1}{2}\log{\frac{\Lambda}{3w^{2}_0}} +\frac{5}{24}\Big\}+S_{\text{corner}}$ containing divergent terms that are $t_0$ independent.
Apart from the $\log{\cosh{\Big[\sqrt{\frac{\Lambda}{3}}t_0\Big]}}$ term the $t_0$ dependent terms of (\ref{totsactiont}) are bounded and monotonic for $t_0>0$. Fourier transforming the amplitude with $\tilde{S} = S - S_{0}$ and employing a similar expansion as (\ref{1p1pole}) reveals the pole again,
\begin{align}\label{3p1pole}
\int^{\infty}_{0}d\,t_{0}\,e^{i (\tilde{S}[t_0] - \omega\,t_0)}&=\int^{\infty}_{0}d\,t_{0}\, e^{i(2\frac{4\pi}{\kappa}\sqrt{\frac{3}{\lambda}}t_{0}-\omega t_0)}+\ldots\nonumber\\&=\frac{i}{\omega -2\frac{4\pi}{\kappa}\sqrt{\frac{3}{\Lambda}}}  +\ldots.
\end{align}

Again we have a pole in the spectral representation at the energy of the static patch. This term is present in  $d+1$ dimensions.

\section{Discussion and Conclusions}
The main implication  of this paper is that there exist transition amplitudes between excited states of supersymmetric flat vacua employed in string theory that possess \dS vacua as resonances. Although we have not mentioned it, a given \dS vacuum contains an exponentially large number of almost degenerate states and in a real quantum theory we would expect a correspondingly dense collection of poles. This is analogous to the idea of a  black hole as a collection of resonances.

None of this should be taken to mean that ordinary scattering amplitudes for finite numbers of particles contain dS. The $|\ein\rangle$ and $|\eout\rangle$ states we are discussing are open (k=-1) FRW cosmologies that contain an infinite number of particles. The particles are uniformly distributed on hyperbolic surfaces and, in particular, there exists an infinite number of particles on $\Sigma$ of Fig. \ref{fig:CDLconstrainedCDL}(left). These excited states manifest as domain walls and dS should be thought of as a resonance between these domain walls. 

We suggest that states of this type form a superselection sector in which the \dS resonances are found. Since these states contain an infinite number of particles but their entropy must not exceed the finite \dS entropy of the causal patch, they must be infinitely fine-tuned. Such states would be the bulk states of FRW/CFT \cite{Freivogel:2004rd,Freivogel:2006xu,Susskind:2007pv,Sekino:2009kv} or similar string theory construction that possesses \dS as an intermediate configuration.

It has been asked how recent work on complexity and relations between geometry and entanglement apply in a cosmological setting.
In $2+1$ dimensions the action calculation when continued to AdS is similar to Wick rotated calculations relating complexity  to action in the AdS BTZ black hole \cite{Brown:2015bva,Brown:2015lvg}. In the continuation $\mathcal{V}$ replaces the Wheeler-DeWitt patch of \cite{Brown:2015bva,Brown:2015lvg}.  In both cases the action grows linearly with time $t_0$, which in the \dS case leads to the resonant pole found; in the AdS version it represents the linear growth in complexity.  It is possible that in cosmology the exponential expansion of space may also represent a growth in complexity; further study in this direction is demanded. A longer paper containing details of the computation will be released concurrently \cite{Maltz:2016max}.

\section{Acknowledgements}
\begin{acknowledgments}
The authors thank Ying Zhao for extremely helpful discussions in the course of this work. Furthermore, the authors thank  Ahmed Almheiri, Dionysios Anninos, Tom Banks, Ning Bao, Adam Brown, David Berenstein, Evan Berkowitz, Raphael Bousso, Ben Frivogel, Ori Ganor, Masanori Hanada, Stefan Leichenauer, Don Page, Micheal Peskin, Douglas Stanford, Raphael Sgier, Yasuhiro Sekino, and Jason Weinberg  for stimulating discussions and comments. J.M. also thanks Timothy Mernard and Nicholas Johnson for stimulating discussions and hospitality during the completion of this work. The work of J.M. is supported by the California Alliance fellowship (NSF Grant No. 32540). Support for the research of L.S. came through NSF Grant No. Phy-1316699 was supported in part by a grant from the John Templeton Foundation and  the Stanford Institute for Theoretical Physics.
\end{acknowledgments}

\bibliographystyle{apsrev4-1}
\bibliography{bibliography}
\end{document}